\pdfoutput=1

\documentclass[11pt]{article}

\newif\iftr

\trtrue

\usepackage[top=1.0in,bottom=1.0in,left=1.0in,right=1.0in]{geometry}

\usepackage{url}
\usepackage{listings}
\newcommand{\var}[1]{\text{\lstinline+#1+}}

\usepackage{changepage}
\usepackage{titlesec}
\usepackage{amsmath}
\usepackage{amssymb}
\usepackage{wrapfig}
\usepackage{tikz}
\usepackage{mathabx}
\usepackage{qtree}
\usepackage{harpoon}
\usepackage{graphicx}
\usepackage{epstopdf}
\usepackage{epsfig}
\usepackage{color}
\usepackage{program}
\usepackage{multicol}
\usepackage{multirow}
\usepackage{proof}
\usepackage{url}
\usepackage{hyperref}
\usepackage{stmaryrd}
\usepackage[textsize=small]{todonotes}
\usepackage{fancyvrb}
\usepackage[normalem]{ulem}

\titlespacing*{\section}
{0pt}{0.3cm}{0.2cm}
\titlespacing*{\subsection}
{0pt}{0.3cm}{0.2cm}

\titlespacing*{\paragraph}
{0pt}{0.2cm}{0.1cm}

\newcommand\Loadedframemethod{default}
\usepackage[framemethod=\Loadedframemethod]{mdframed}

\usepackage{listings}
\usepackage{lstlinebgrd}
\let\origthelstnumber\thelstnumber
\makeatletter
\newcommand*\Suppressnumber{%
  \lst@AddToHook{OnNewLine}{%
    \let\thelstnumber\relax%
     \advance\c@lstnumber-\@ne\relax%
    }%
}

\newcommand*\Reactivatenumber{%
  \lst@AddToHook{OnNewLine}{%
   \let\thelstnumber\origthelstnumber%
   \advance\c@lstnumber\@ne\relax}%
}

\lstdefinestyle{nonumbers}
               {numbers=none}

\usepackage{algorithm}
\usepackage{algpseudocode}
\algrenewcommand{\algorithmiccomment}[1]{{\small\hfill$\triangleright$ #1}}
\algrenewcommand\algorithmicindent{1em}
\algnewcommand\algorithmicswitch{\textbf{switch}}
\algnewcommand\algorithmiccase{\textbf{case}}
\algnewcommand\algorithmicassert{\texttt{assert}}
\algnewcommand\Assert[1]{\State \algorithmicassert(#1)}%
\algdef{SE}[SWITCH]{Switch}{EndSwitch}[1]{\algorithmicswitch\ #1\ \algorithmicdo}{\algorithmicend\ \algorithmicswitch}%
\algdef{SE}[CASE]{Case}{EndCase}[1]{\algorithmiccase\ #1}{\algorithmicend\ \algorithmiccase}%
\algtext*{EndSwitch}%
\algtext*{EndCase}%
\algtext*{EndFor}%
\algtext*{EndIf}%
\newenvironment{itemize*}%
  {\begin{itemize}%
    \setlength{\itemsep}{0.0in}%
    \setlength{\topsep}{0.0in}%
    \setlength{\parskip}{0.0in}}%
  {\end{itemize}}
\newenvironment{enumerate*}%
  {\begin{enumerate}%
    \setlength{\itemsep}{0.0in}%
    \setlength{\topsep}{0.0in}%
    \setlength{\parskip}{0.0in}}%
  {\end{enumerate}}

\definecolor{light-gray}{gray}{0.85}

\newcommand\red[1]{{\color{red} #1}}

\newcommand\ignore[1]{}

\newcommand\FULL[1]{\red{(omitted content)}}

\renewcommand{\paragraph}[1]{\vspace{.6em}\noindent\textbf{#1}\hspace{.5em}}

\iftr
\title{Adding Concurrency to Smart Contracts}
\else
\title{Adding Concurrency to Smart Contracts \\
  {\large (Regular Submission)}}
\fi
\author{
  Thomas Dickerson\\
  Brown University\\
  \texttt{thomas\_dickerson@brown.edu}
  \and
  Paul Gazzillo\\
  Yale University\\
  \texttt{paul.gazzillo@yale.edu}
  \and
  Maurice Herlihy\\
  Brown University\\
  \texttt{maurice\_herlihy@brown.edu}
  \and
  Eric Koskinen\\
  Yale University\\
  \texttt{eric.koskinen@yale.edu}
}

\date{}
\begin{document}

\newcommand\BallotcontentionMiner{1.57}
\newcommand\BallotcontentionValidator{1.73}
\newcommand\BallotblocksizeMiner{1.35}
\newcommand\BallotblocksizeValidator{1.58}
\newcommand\SimpleAuctioncontentionMiner{1.23}
\newcommand\SimpleAuctioncontentionValidator{1.35}
\newcommand\SimpleAuctionblocksizeMiner{1.58}
\newcommand\SimpleAuctionblocksizeValidator{1.60}
\newcommand\EtherDoccontentionMiner{0.78}
\newcommand\EtherDoccontentionValidator{2.04}
\newcommand\EtherDocblocksizeMiner{1.09}
\newcommand\EtherDocblocksizeValidator{1.75}
\newcommand\MixedcontentionMiner{1.57}
\newcommand\MixedcontentionValidator{1.86}
\newcommand\MixedblocksizeMiner{1.45}
\newcommand\MixedblocksizeValidator{1.64}
\newcommand\overallMinerSpeedup{1.33}
\newcommand\overallValidatorSpeedup{1.69}

\iftr
\maketitle 
\begin{abstract}
  Modern cryptocurrency systems, such as Ethereum,
permit complex financial transactions through scripts called \emph{smart contracts}.
These smart contracts are executed many, many times,
always without real concurrency.
First, all smart contracts are serially executed by \emph{miners} 
before appending them to the blockchain.
Later, those contracts are serially re-executed by \emph{validators}
to verify that the smart contracts were executed correctly by miners.
\iftr

\fi
Serial execution limits system throughput
and fails to exploit today's concurrent multicore and cluster architectures.
Nevertheless, serial execution appears to be required:
contracts share state,
and contract programming languages have a serial semantics.
\iftr

\fi
This paper presents a novel way to permit miners and validators to
execute smart contracts in parallel,
based on techniques adapted from software transactional memory.
Miners execute smart contracts speculatively in parallel,
allowing non-conflicting contracts to proceed concurrently,
and ``discovering'' a serializable concurrent schedule for a block's transactions,
This schedule is captured and encoded as a deterministic \emph{fork-join} program used
by validators to re-execute the miner's parallel schedule
deterministically but concurrently.
\iftr

\fi
Smart contract benchmarks run on a JVM with ScalaSTM
show that a speedup of \overallMinerSpeedup{}x can be obtained for
miners and \overallValidatorSpeedup{}x for validators with just three concurrent threads.

\end{abstract}
\fi

\section{Introduction}

Cryptocurrencies such as Bitcoin~\cite{bitcoin} or
Ethereum~\cite{ethereum} are very much in the news.
Each is an instance of a \emph{distributed ledger}:
a publicly-readable tamper-proof record of a sequence of events.
Simplifying somewhat, early distributed ledgers, such as Bitcoin's, work like this:
\emph{clients} send
\emph{transactions}\footnote{Following blockchain terminology,
  a transaction is a payment or set of payments,
  not an atomic unit of synchronization as in databases or transactional memory.}
to \emph{miners},
who package the transactions into \emph{blocks}.
Miners repeatedly \emph{propose} new blocks to be applied to the ledger,
and follow a global consensus protocol to agree on which blocks are chosen.
Each block contains a cryptographic hash of the previous block,
making it difficult to tamper with the ledger.
The resulting distributed data structure, called a \emph{blockchain},
defines the sequence of transactions that constitutes the distributed
ledger\footnote{This description omits many important issues,
such as incentives, forking, and fork resolution.}

Modern blockchain systems often interpose an additional software layer between
clients and the blockchain.
Client requests are directed to scripts, called \emph{smart contracts},
that perform the logic needed to provide a complex service,
such as managing state, enforcing governance, or checking credentials.
Smart contracts can take many forms,
but here we will use (a simplified form of) the Ethereum model~\cite{ethereum}.

A smart contract resembles an object in a programming language.
It manages long-lived \emph{state}, which is encoded in the blockchain.
The state is manipulated by a set of \emph{functions},
analogous to \emph{methods} in many programming languages.
Functions can be called either directly by clients or indirectly by other
smart contracts.
Smart contract languages are typically Turing-complete.
To ensure that function calls terminate,
the client is charged for each computational step in a function call.
If the charge exceeds what the client is willing to pay,
the computation is terminated and rolled back.

When and where is smart contract code executed?
There are two distinct circumstances.
Each smart contract is first executed by one or more \emph{miners},
nodes that repeatedly propose new blocks to append to the blockchain.
When a miner creates a block,
it selects a sequence of user requests
and executes the associated smart contract code for each Ethereum transaction
in sequence,
transforming the old contract state into a new state.
It then records both the sequence of transactions and the new state in
the block, and proposes it for inclusion in the blockchain.

Later, when the block has been appended to the blockchain,
each smart contract is repeatedly re-executed by \emph{validators}:
nodes that reconstruct (and check) the current blockchain state.
As a validator acquires each successive block,
it replays each of the transactions' contract codes to check that the
block's initial and final states match.
Each miner validates blocks proposed by other miners,
and older block are validated by newly-joined miners,
or by clients querying the contract state.
Code executions for validation vastly exceed code executions for mining.

Existing smart contract designs limit throughput because they admit no concurrency.
When a miner creates a block,
it assembles a sequence of transactions,
and computes a tentative new state by executing  those transactions' smart contracts
serially, in the order they occur in the block.
A miner cannot simply execute these contracts in parallel,
because they may perform conflicting accesses to shared data,
and an arbitrary interleaving could produce an inconsistent final state.
For Bitcoin transactions, it is easy to tell in advance when two transaction conflict,
because input and output data are statically declared.
For smart contracts, by contrast,
it is impossible to tell in advance whether two contract codes conflict,
because the contract language is Turing-complete.

Miners are rewarded for each block they successfully append to the blockchain,
so they have a strong incentive to increase throughput by parallelizing
smart contract executions.
We propose to allow miners to execute contract codes in parallel by adapting
techniques from Software Transactional Memory (STM)~\cite{Herlihy:2003:STM:872035.872048}:
treating each invocation as a speculative atomic action.
Data conflicts, detected at run-time,
are resolved by delaying or rolling back some conflicting invocations.
Treating smart contract invocations as speculative atomic actions
dynamically ``discovers a \emph{serializable} concurrent schedule,
producing the same final state as a serial schedule
where the contract functions were executed in some one-at-a-time order.

But what about later validators?
Existing STM systems are \emph{non-deterministic}:
if a later validator simply mimics the miner by
re-running the same mix of speculative transactions,
it may produce a different serialization order and a
different final state,
causing validation to fail incorrectly.
Treating contract invocations as speculative transactions
improves miners' throughput,
but fails to support deterministic re-execution as required by validators.

Notice, however,
that the miner has already ``discovered'' a serializable concurrent schedule for those transactions.
We propose a novel scheme where the miner records that successful schedule,
along with the final state,
allowing later validators to replay that same schedule in a concurrent but deterministic way.
Deterministic replay avoids many of the the miner's original synchronization costs,
such as conflict detection and roll-back.
Over time,
parallel validation would be a significant benefit because validators perform
the vast majority of contract executions.
Naturally,
the validator must be able to check that the proposed schedule
really is serializable.

This paper makes the following contributions.
\begin{itemize*}
\item A way for miners to speculatively execute smart contracts in parallel.
  We adapt techniques from \emph{transactional boosting}~\cite{HerlihyK2008}
  to permit non-conflicting smart contracts to execute concurrently.

\item A way for miners to capture the resulting parallel execution in the form of a
  \emph{fork-join}~\cite{BlumofeJKLRZ1995} schedule to be executed by validators,
  deterministically, verifiably, and in parallel.

\item A prototype implementation, built on the Java virtual machine and ScalaSTM~\cite{scalastm}.
  An evaluation using smart contract examples drawn from the Solidity documentation
  yields an overall speedup of \overallMinerSpeedup{}x for miners,
  and \overallValidatorSpeedup{}x for validators with three concurrent threads of execution.
\end{itemize*}

\section{Blockchains and Smart Contracts}
\label{sec:model}

\begin{wrapfigure}[18]{r}{3.7in}
\begin{lstlisting}[caption={Part of the Ballot contract.}, label={lst:ballot}]
contract Ballot {`\label{ln:defBallot}`
  mapping(address => Voter) public voters; `\label{ln:defVoters}`
  // more state definitions
  function vote(uint proposal) {`\label{ln:vote}`
    Voter sender = voters[msg.sender];
    if (sender.voted)
      throw;
    sender.voted = true;
    sender.vote = proposal;
    proposals[proposal].voteCount += sender.weight;
  }
  // more operation definitions
}
\end{lstlisting}
\end{wrapfigure}

In Bitcoin and similar systems,
transactions typically have a simple structure,
distributing the balances from a set of input accounts to a set of
newly-created output accounts.
In Blockchains such as Ethereum, however,
each block also includes an explicit \emph{state} capturing the cumulative
effect of transactions in prior blocks.
Transactions are expressed as executable code,
often called \emph{smart contracts},
that modifies that state.
Ethereum blocks thus contain both transactions' smart contracts
and the final state produced by executing those contacts.

The contracts themselves are stored in the blockchain as byte-code instructions for 
the Ethereum virtual machine (EVM).
Several higher-level languages exist for writing smart contracts.
Here, we describe
smart contracts as expressed in the Solidity language~\cite{solidity}.

\newcommand\vvv[1]{\lstinline`#1`}

Listing~\ref{lst:ballot} is part of the source code for an example smart
contract that implements a ballot box~\cite{solidityexamples}.
The owner
initializes the contract with a list of proposals and gives
the right to vote to a set of Ethereum addresses.
Voters cast their votes for a particular proposal,
which they may do only once.
Alternatively, voters may delegate their vote.
The \vvv{contract} keyword declares the smart contract (Line~\ref{ln:defBallot}).

The contract's persistent state is recorded in \emph{state variables}. For \vvv{Ballot}, the persistent state includes fields of scalar type such as the owner (omitted for lack of space).
State variables such as \vvv{voters} (declared on
Line~\ref{ln:defVoters}) can also use the built-in Solidity type
\vvv{mapping} which, in this case, associates each voter's
\vvv{address} with a \vvv{Voter} data structure (declaration omitted
for brevity). 
The keys in this mapping are of built-in type \vvv{address},
which uniquely identifies Ethereum accounts (clients or other contracts).
These state variables are the persistent state of the contract.

Line~\ref{ln:vote} declares contract \emph{function}, \vvv{vote}, to
cast a vote for the given proposal. 
Within a function there are transient \emph{memory} and \emph{stack} areas
such as \vvv{sender}.
The function \vvv{vote} first recovers the \vvv{Voter} data from the
contract's state by indexing into the \vvv{voters} mapping using the
sender's address \vvv{msg.sender}.
The \vvv{msg} variable is a global variable containing data about the
contract's current invocation.
Next, the \vvv{sender.vote} flag is checked to prevent multiple votes.
Note that sequential execution is critical:
if this code were na{\"\i}vely run in parallel,
it would be vulnerable to a race condition permitting double voting.
Ethereum contract functions can be aborted at any time via
\vvv{throw}, as seen here when a voter is detected attempting
to vote twice.
The \vvv{throw} statement causes the contract's transient state and
tentative storage changes to be discarded.
Finally, this \vvv{Ballot} contract also provides functions to register voters, delegate
one's vote, and compute the winning proposal.  The complete Ballot
example is shown in Appendix~\ref{apx:contract}.

\paragraph{Execution Model: Miners and Validators.}
When a miner prepares a block for inclusion in the blockchain,
it starts with the ledger state as of the chain's most recent block.
The miner selects a sequence of new transactions,
records them in the new block,
and executes those their contracts, one at a time,
to compute the new block's state.
The miner then participates in a consensus protocol to decide
whether this new block will be appended to the blockchain.

To ensure that each transaction terminates in a reasonable number
of steps, each call to contract bytecode comes with an explicit limit on
the number of virtual machine steps that a call can take.
(In Ethereum,
these steps are measured in gas and clients pay a fee to the miner that
successfully appends that transaction's block to the blockchain.)

After a block has been successfully appended to the blockchain,
that block's transactions are sequentially re-executed
\emph{by every node in the network}
to check that the block's state transition was computed honestly and correctly.
(Smart contract transactions are deterministic,
so each re-execution yields the same results as the original.)
These \emph{validator} nodes do not receive fees for re-execution.

To summarize,
a transaction is executed in two contexts:
once by miners before attempting to append a block to the blockchain,
and many times afterward by validators checking that each block in the blockchain
is honest.
In both contexts, each block's transactions are executed sequentially in block-order.

\section{Speculative Smart Contracts}
\label{sec:boosting}
This section discusses how miners can execute contract codes
concurrently.
Concurrency for validators is addressed in the next section.

Smart contract semantics is \emph{sequential}:
each miner has a single thread of control that executes one EVM instruction at a time.
The miner executes each of the block's contracts in sequence.
One contract can call another contract's functions,
causing control to pass from the first contract code to the second, and back again.
(Indeed, misuse of this control structure has been the source of
well-known security breaches~\cite{theDao}.)
Clearly,
even sequential smart contracts must be written with care,
and introducing explicit concurrency to contract programming languages
would only make the situation worse.
We conclude that concurrent smart contract executions must be
\emph{serializable}, indistinguishable, except for execution time,
from a sequential execution.

There are several obstacles to running contracts in parallel.
First, smart contract codes read and modify shared storage,
so it is essential to ensure that concurrent contract code executions do not result
in inconsistent storage states.
Second,
smart contract languages are Turing-complete,
and therefore it is impossible to determine statically whether
contracts have a data conflict. 

We propose that miners execute contract codes as \emph{speculative actions}.
A miner schedules multiple concurrent contracts to run in parallel.
Contracts' data structures are instrumented
to detect synchronization conflicts at run-time,
in much the same way as mechanisms like transactional boosting~\cite{HerlihyK2008}.
If one speculative contract execution conflicts with another,
the conflict is resolved either by delaying one contract until another completes,
or by rolling back and restarting some of the contracts.
When a speculative action completes successfully,
it is said to \emph{commit}, and otherwise it \emph{aborts}.

\paragraph{Storage Operations.}
We assume that, as in Solidity,
state variables are restricted to predefined types such as scalars,
structures, enumerations, arrays, and mappings.
A \emph{storage operation} is a primitive operation on a state
variable.
For example, binding a key to a value in a mapping,
or reading from an variable or an array are storage operations.
Two storage operations \emph{commute} if executing them in either
order yields the same result values and the same storage state.
For example, in the \vvv{address}-to-\vvv{Voter} \vvv{Ballot} mapping in
Listing~\ref{lst:ballot}, binding Alice's address to a vote of
42 commutes with binding Bob's address to a vote of 17, but does not
commute when deleting Alice's vote.
An \emph{inverse} for a storage operation is another operation
that undoes its effects.
For example,
the inverse of assigning to a variable is restoring its prior value,
and the inverse of adding a new key-value pair to a mapping is to
remove that binding, and so on.
We assume the run-time system provides all storage operations with inverses.

The virtual machine is in charge of managing concurrency for state
variables such as mappings and arrays.
Speculation is controlled by two run-time mechanisms,
invisible to the programmer,
and managed by the virtual machine:
\emph{abstract locks}, and \emph{inverse logs}.

Each storage operation has an associated abstract lock.
The rule for assigning abstract locks to operations is simple:
if two storage operations map to distinct abstract locks,
then they must commute.
Before a thread can execute a storage operation,
it must acquire the associated abstract lock.
The thread is delayed while that lock is held by another thread%
\footnote{For ease of exposition,
abstract locks are mutually exclusive,
although it is not hard to accommodate shared and exclusive modes.}.
Once the lock is acquired, 
the thread records an \emph{inverse operation} in a log,
and proceeds with the operation.

If the action commits,
its abstract locks are released and its log is discarded.
If the action aborts,
the inverse log is replayed,
most recent operation first,
to undo the effects of that speculative action.
When the replay is complete, the action's abstract locks are released.

The advantage of combining abstract locks with inverse logs is that
the virtual machine can support very fine-grained concurrency.
A more traditional implementation of speculative actions might
associate locks with memory regions such as cache lines or pages,
and keep track of old and versions of those regions for recovery.
Such a coarse-grained approach could lead to many false conflicts,
where operations that commute in a semantic sense are treated as
conflicting because they access overlapping memory regions.
In the next section,
we will see how to use abstract locks to speed up verifiers.
  
When one smart contract calls another,
the run-time system creates a \emph{nested} speculative action,
which can commit or abort independently of its parent.
A nested speculative action inherits the abstract locks held by its parent,
and it creates its own inverse log.
If the nested action commits,
any abstract locks it acquired are passed to its parent,
and its inverse log is appended to its parent's log.
If the nested action aborts,
its inverse log is replayed to undo its effects,
and any abstract locks it acquired are released.
Aborting a child action does not abort the parent,
but a child action's effects become permanent only when the parent commits.
The abstract locking mechanism also detects and resolves deadlocks,
which are expected to be rare.

The scheme described here is \emph{eager},
acquiring locks, applying operations, and recording inverses.
An alternative \emph{lazy} implementation could buffer changes to
a contract's storage, applying them only on commit.

A miner's incentive to perform speculative concurrent execution
is the possibility of increased throughput,
and hence a competitive advantage against other miners.
Of course,
the miner undertakes a risk that synchronization conflicts among
contracts will cause some contracts to be rolled back and re-executed,
possibly delaying block construction,
and forcing the miner to re-execute code not compensated by client fees.
Nevertheless,
the experimental results reported below suggest that even a small
degree of concurrent speculative execution pays off,
even in the face of moderate data conflicts.

\section{Concurrent Validation}
\label{sec:replay}

The speculative techniques proposed above for miners are no help for validators.
Here is the problem:
miners use speculation to discover a concurrent schedule for a block's transactions,
a schedule equivalent to some sequential schedule, except faster.
That schedule is constructed non-deterministically,
depending on the order in which threads acquired abstract locks.
To check that the block's miner was honest,
validators need to reconstruct the same (or an equivalent) schedule
chosen by the miner.

Validators need a way to deterministically reproduce the miner's concurrent schedule.
To this end, we extend abstract locks to track dependencies,
that is, who passed which abstract locks to whom.
Each speculative lock includes a \emph{use counter} that keeps track
of the number of times it has been released by a committing action during the
construction of the current block.
When a miner starts a block, it sets these counters to zero.

When a speculative action commits,
it increments the counters for each of the locks it holds,
and then it registers a \emph{lock profile} with the VM recording the
abstract locks and their counter values. 

When all the actions have committed,
it is possible to reconstruct their common schedule by comparing their
lock profiles.
For example, consider three committed speculative actions, $A$, $B$,
and $C$.
If $A$ and $B$ have no abstract locks in common, they can run concurrently.
If an abstract lock has counter value $1$ in $A$'s profile
and $2$ in $C$'s profile,
then $C$ must be scheduled after $A$.

\iftr
\begin{algorithm}
%% \linespread{1.10}\normalsize
\caption{$\textsc{MineInParallel}(T)$ - Mine in parallel}
\label{alg:mine}
\begin{algorithmic}[1]
  \Require A set of contract transactions $T$
  \Ensure A serial order $S$ of transactions and a happens-before graph $H$ of the locking schedule
  \Function{MineInParallel}{$B$}
  \State {Initialize log $L$ for recording locking operations}
  \State {Execute all transactions $t \in T$ in parallel, recording locking activity in $L$}
  \State {Generate happens-before graph $H$ from $L$}
  \State {Create the serial ordering $S$ via a topological sort of $H$}
  \State {\Return $(S, H)$}
  \EndFunction
\end{algorithmic}
\end{algorithm}

\begin{algorithm}
%% \linespread{1.10}\normalsize
\caption{$\textsc{ConstructValidator}(S, H)$ - Construct a parallel validator}
\label{alg:validate}
\begin{algorithmic}[1]
  \Require The serial ordering $S$ and happens-before graph $H$ from the miner
  \Ensure A set of fork-join tasks ensuring parallel execution according to the happens-before graph
  \Function{ConstructValidator}{$B$}
  \State {Initialize a mapping $F$ from each transaction $t$ to its fork-join task $f$}
  \State {Create the happens-after graph $H'$ by reversing the edges of $H$}
  \ForAll {$t \in S$}
  \State {$B \gets$ all transactions $u \in H'$ that happen immediately before $t$, i.e., its outedges}
  \State {Create a fork-join task $f$ for $t$ that first joins with all tasks in $B$, i.e.,}
\begin{lstlisting}[style=nonumbers]
     $f \gets $ for ($b$ in $B$) { $F$.get($b$).join() } execute($t$)
\end{lstlisting}
  \State {Save the new fork-join task in $F$, i.e., $F.\texttt{put}(t, f)$}
  \EndFor
  \State \Return the value set of $F$, the fork-join tasks
  \EndFunction
\end{algorithmic}
\end{algorithm}

A miner includes these profiles in the blockchain along with usual information.
From this profile information,
validators can construct a \emph{fork-join} program that
deterministically reproduces the miner's original, speculative schedule.
Algorithm~\ref{alg:mine} provides a high-level sketch of the operation
of the miner.  By logging the locking schedule during parallel
execution, the miner generates a happens-before graph of transactions
according to the order in which they acquire locks and commit.  A
valid serial history is produced from a topological sort of this
graph.
Algorithm~\ref{alg:validate} constructs
the validator by scanning through the list of actions as they
appear in the serial history.  A fork-join task is created for each
action and stored for lookup by its identifier.  Each
task will first lookup and join any tasks that must precede it
according to the locking schedule before executing the action
itself.
\else
A miner includes these profiles in the blockchain along with usual information.
From this profile information,
validators can construct a \emph{fork-join} program that
deterministically reproduces the miner's original, speculative schedule.
For lack of space, the algorithms for generating the schedule from the
miner and constructing the validator are given in
Appendix~\ref{apx:algorithms}.  Intuitively, the algorithm constructs
the validator by scanning through the list of actions as they
appear in the serial history.  A fork-join task is created for each
action and stored for lookup by its identifier.  Each
task will first lookup and join any tasks that must precede it
according to the locking schedule before executing the action
itself.
\fi

The resulting fork-join program is not speculative,
nor does it require inter-thread synchronization other than forks and joins.
The validator is not required to match the miner's level of parallelism:
using a work-stealing scheduler~\cite{BlumofeJKLRZ1995},
the validator can exploit whatever degree of parallelism it has available.
The validator does not need abstract locks,
dynamic conflict detection,
or the ability to roll back speculative actions,
because the fork-join structure ensures that conflicting actions never
execute concurrently.

To check that the miner's proposed schedule is correct,
the validator's virtual machine records a trace of
the abstract locks each thread would have acquired,
had it been executing speculatively.
This trace is thread-local,
requiring no expensive inter-thread synchronization.
At the end of the execution,
the validator's VM compares the traces it generated with the lock
profiles provided by the miner.
If they differ, the block is rejected.

What is a miner's incentive to be honest about its fork-join schedule?
A miner who publishes an incorrect schedule will be detected and its
block rejected,
but a miner may be tempted to publish a correct sequential schedule equivalent to,
but slower than its actual parallel schedule,
with the goal of slowing down verification by competing miners.
Perhaps the simplest way to provide an incentive is to reward
miners more for publishing highly parallel schedules
(for example, as measured by critical path length).
This reward could be part of a static ``coinbase'' transaction that creates currency,
or client fees could be discounted for less parallel schedules.
Because fork-join schedules are published in the blockchain,
their degree of parallelism is easily evaluated.
Naturally, such rewards must be calibrated to produce desired effects,
a subject beyond the scope of this paper.

\section{Correctness}
Concurrent calls to smart contract functions might leave persistent storage
in an inconsistent state not possible after a serial execution.
Instead, we must show that every concurrent execution permitted by our
proposal is equivalent to some sequential execution.
Because miners are free to choose the order in which contracts appear in a block,
any sequential execution will do.

Our argument builds on the prior proofs
that transactional boosting is serializable~\cite{KPH:POPL2010,KP:PLDI2015,HerlihyK2008}.
A given execution of a contract's function involves
a sequence of operations on storage objects.
(The Ethereum gas restriction ensures this sequence is finite.)
Recall that if two storage operations map to distinct abstract locks,
then they commute.
If another thread executes another sequence of operations,
then if there are two operations that do not commute,
then both threads will try to acquire the same lock,
and one will be delayed until the other completes.
(Deadlocks are detected and resolved by aborting one execution.)
As proved elsewhere~\cite{KPH:POPL2010,KP:PLDI2015,HerlihyK2008},
the result is a serializable execution\footnote{
  Because speculative executions take place entirely within a virtual machine,
  opacity~\cite{opacity} is not an issue.}.

We cannot guarantee that the schedule published by the miner is the
same one that it executed,
but we can guarantee the two are equivalent to a common sequential history.
Validators replay the concurrent schedule published by the miner,
and will detect if the schedule produces a final state
different from the one recorded in the block,
or if the schedule has a data race
(an unsynchronized concurrent access).

\section{Implementation}
Because the EVM is not multithreaded,
our prototype uses the Java Virtual Machine (JVM).
Speculative actions are executed by the Scala Software Transactional
Memory Library (ScalaSTM~\cite{scalastm}). 

Examples of smart contracts were translated from Solidity into Scala,
then modified to use the concurrency libraries.
Each function from the Solidity contract is turned into a speculative
transaction by wrapping its contents with a ScalaSTM \vvv{atomic}
section.
Solidity \vvv{mapping} objects are implemented as boosted hashtables,
where key values are used to index abstract locks.
Additionally, solidity \vvv{struct} types were translated into immutable
case classes.
Methods take a \var{msg} field to emulate 
Solidity contracts' global state,
which includes details of the transaction,
addresses of participants, and so on.
Scalar fields are implemented as a single a boosted mapping.

The Solidity \var{throw} operation,
which explicitly rolls back a contract execution,
is emulated by throwing a Java runtime exception caught by the miner.

In our prototype,
abstract locks are implemented via interfaces exported by ScalaSTM,
relying on ScalaSTM's native deadlock detection and resolution mechanisms.

\subsection{Miners and Validators}

Miners manage concurrency using Java's \vvv{ExecutorService}.
This class provides a pool of threads and runs a collection of
\vvv{callable} objects in parallel.
A block of transactions in Ethereum is implemented as a set of
\vvv{callable} objects passed to the thread pool.
To generate locking profiles from the parallel execution,
we instrument smart contracts to log when atomic sections start and end,
as well as calls to boosted operations.
From the log, we can encode the locking schedule as a
happens-before graph for the validator.
The validator transforms this happens-before graph into a fork-join program.
Each transaction from the block is a fork-join task that first joins with
all tasks according to its in-edges on the happens-before graph.

\newcommand\benchstamp{20170208-105017}
\newcommand\benchstampt[1]{exp/\benchstamp-#1}

\section{Experimental Evaluation}
Our goal is to improve throughput for miners and validators by
allowing unrelated contracts to execute in parallel.
To evaluate this approach,
we created a series of benchmarks for sample
contracts that vary the number of transactions and their degree of conflict.
These benchmarks are conservative,
operating on only one or a few contracts at a time and permitting higher
degrees of data conflict than one would expect in practice.

Our experiments are designed to answer two questions.
(1) For a given amount of data conflict, how does speedup
change over increasing transactions?  We expect to see more speedup as
the number of transactions increases, limited by the number of cores
available on the underlying hardware.
(2) How does the speedup change as data conflict increases?
For low data conflict,
we expect our parallel miner to perform better than serial.
But as data conflict increases,
we expect a drop-off in speedup,
limited by core availability.

\subsection{Benchmarks}

There are four benchmarks, one for each of the example contracts we
implemented, Ballot, SimpleAuction, and EtherDoc, as well as the Mixed
benchmark containing transactions from all.  For each benchmark, our
implementation is evaluated on blocks containing between 10 and 400
transactions with 15\% data conflict, as well as blocks containing 200
transactions with data conflict percentages ranging from 0\% to 100\%
data conflict.
The data conflict percentage is defined to be the percentage
of transactions that contend with at least one other transaction for
shared data.  As we will see, the impact of data conflict on speedup
depends on the contract implementation.

These benchmarks are conservative.  For all
benchmarks besides Mixed, the entire block operates on the same
contract, calling only one or two methods.  In reality, mined blocks
contained transactions on unrelated contracts and accounts.  While the
theoretical maximum number of transactions per block is currently
around 200\footnote{A transaction costs 21,000 gas plus the gas for
the computation~\cite{wood}.  The gas limit the block 3,110,235
(latest as of writing) was 4,005,875, a maximum close to 200.}, we
test a wide range from 10 to 400.  The maximum increases and decreases
over time, as determined by miner
preference~\cite{ethereumdesign}.
In practice, the number of transactions can be far fewer per block,
e.g., when there are costly transactions.  For testing speedup over
number of transactions, we fix the data conflict rate at 15\%, though we
expect that blocks in practice rarely have very much internal data conflict.
While we did not measure data conflict in the existing blockchain,
our approach implemented in EVM could be used to collect such data on
an existing blockchain.  For testing speedup as data conflict increases,
we fix the number of transactions per block to 200, the current
theoretical maximum.

\paragraph{Ballot.}
This contract is an example voting application from the Solidity
documentation~\cite{solidityexamples}
and is described in Section~\ref{sec:model}.  For all benchmarks, the
contract is put into an initial state where voters are already
registered.  All block transactions for this benchmark are requests to
vote on the same proposal.  To add data conflict, some voters attempt to
double-vote, creating two transactions that contend for the same voter
data.  100\% data conflict occurs when all voters attempt to vote twice.

\paragraph{SimpleAuction.}
This contract, also from the Solidity
documentation~\cite{solidityexamples}
implements an auction. There is a single owner who initiates the
auction, while any participant can place bids with the \texttt{bid()}
method. A \vvv{mapping} tracks how much money needs to be returned to which
bidder once the auction is over.  Bidders can then \texttt{withdraw()}
their money.  For the benchmarks, the contract state is initialized by
several bidders entering a bid.  The block consists of transactions
that withdraw these bids.  Data Conflict is added by including new
bidders who call \texttt{bidPlusOne()} to read and increase the
highest bid.  The rate of data Conflict depends on how many bidders are bidding at
the same time, thus accessing the same highest bidder. 100\% data conflict
happens when all transactions are \texttt{bidsPlusOne()} bids.

\paragraph{EtherDoc.}
EtherDoc\footnote{\url{https://github.com/maran/notareth}}
is a ``Proof of Existence'' decentralized application (DAPP) that
tracks per-document metadata including hashcode owner.
It permits new document creation,
metadata retrieval, and ownership transfer.  For the benchmarks, the
contract is initialized with a number of documents and owners.
Transactions consist of owners checking the existence of the document
by hashcode.  Data Conflict is added by including transactions that
transfer ownership to the contract creator.  As with SimpleAuction,
all contending transactions touch the same shared data, so we expect a
faster drop-off in speedup with increased data conflict than Ballot.
100\% data conflict happens when all transactions are transfers.

\paragraph{Mixed.}
This benchmark combines transactions on the above smart contracts in
equal proportions, and data conflict is added the same way in equal
proportions from their corresponding benchmarks.

\subsection{Results}

We ran our experiments on a 4-core 3.07GHz Intel Xeon W3550 with 12 GB
of memory running Ubuntu 16.  All of our experiments run on the Java
Virtual Machine (JVM) with JIT compilation disabled.  Parallel mining
and validation are run with a fixed pool of three threads, leaving one
core available for garbage collection and other system
processes/threads.

For each benchmark, blocks were generated for each combination of the
number of transactions and data conflict percentage.  Each block is run
on the parallel miner, the validator, and a serial miner that runs the
block without parallelization.  The serial results serve as the
baseline for computing speedup.  The running time is collected five
times and the mean and standard deviation are measured.  All runs are
given three warm-up runs per collection.

\begin{figure*}
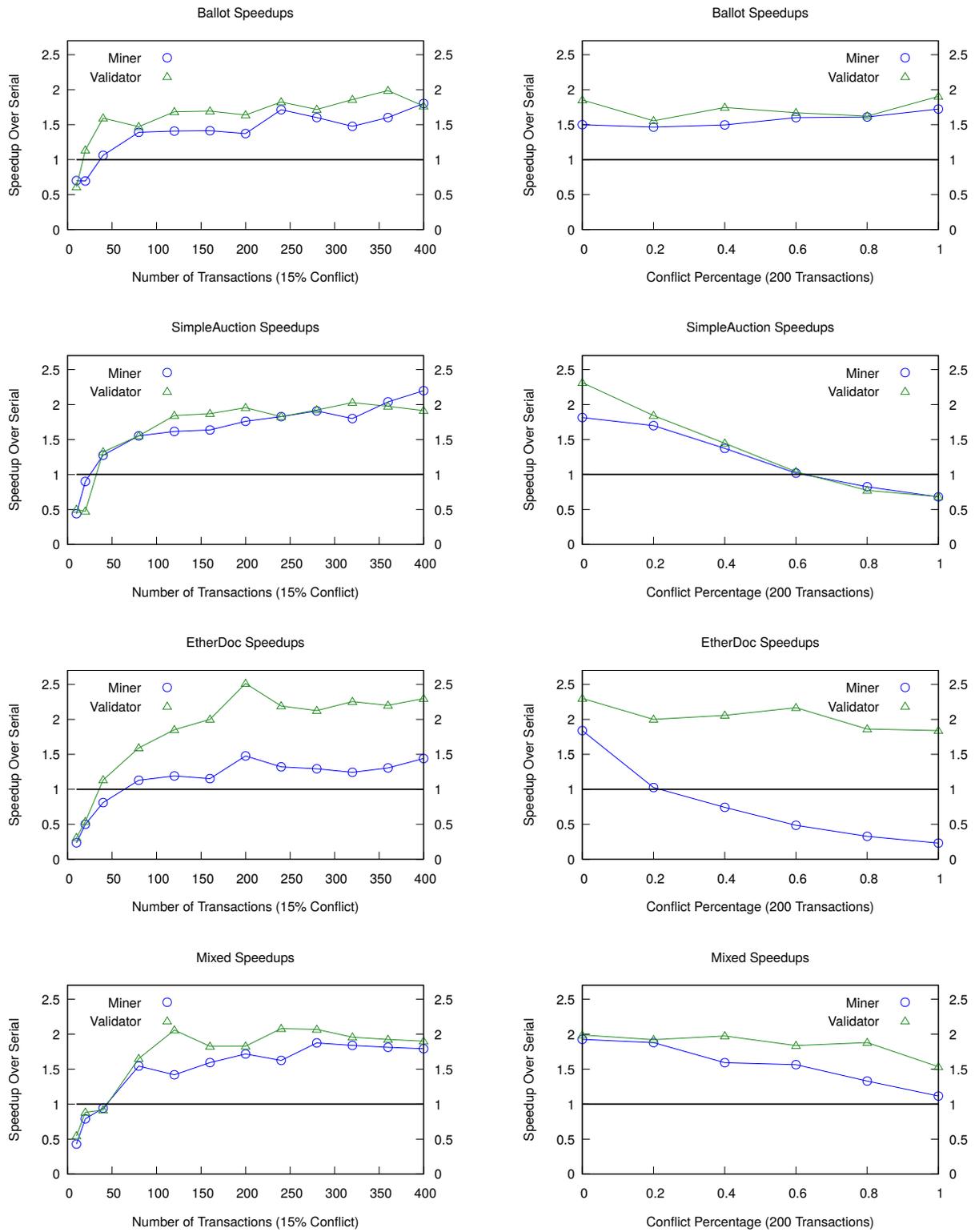
\centering
\begin{tabular}{cc}
\iftr
\includegraphics[width=3.2in]{\benchstampt{Ballot-blocksize-speedup.pdf}} &
\includegraphics[width=3.2in]{\benchstampt{Ballot-contention-speedup.pdf}}\\
\includegraphics[width=3.2in]{\benchstampt{SimpleAuction-blocksize-speedup.pdf}} &
\includegraphics[width=3.2in]{\benchstampt{SimpleAuction-contention-speedup.pdf}}\\
\includegraphics[width=3.2in]{\benchstampt{EtherDoc-blocksize-speedup.pdf}} &
\includegraphics[width=3.2in]{\benchstampt{EtherDoc-contention-speedup.pdf}}\\
\fi
\includegraphics[width=3.2in]{\benchstampt{Mixed-blocksize-speedup.pdf}} &
\includegraphics[width=3.2in]{\benchstampt{Mixed-contention-speedup.pdf}}\\
\end{tabular}
\caption{\label{fig:speedup} The speedup of the miner and validator versus serial mining for
\iftr
each
\else
the Mixed
\fi
benchmark.  The left chart is the speed up as block size increases, while the right is the speed up as data conflict increases.}
\end{figure*}

\iftr
Figure~\ref{fig:speedup} shows the speedup of the parallel miner and
validator relative to the serial miner for all.  (The running times
with mean and standard deviation can be found in
Appendix~\ref{apx:dev}.)
\else
Figure~\ref{fig:speedup} shows the speedup of the parallel miner and
validator relative to the serial miner for the Mixed benchmarks
combining transactions from all smart contracts.  (The speedup charts
for all benchmarks and their running times with mean and standard
deviation can be found in Appendices~\ref{apx:speedups}
and~\ref{apx:dev}.)
\fi
The left chart plots the speedup over the
number of transactions in the block at a fixed data conflict percentage
of 15\%.  The speedup for all benchmarks follows roughly the same
pattern.  For low numbers of transactions, there is no speedup and
even some slowdown.  This is likely due to data conflict as well as the
overhead of multithreading.  For over around 50 transactions, there is
a speedup that increases to about 2x, in line with expectations from a
thread pool of size three.  EtherDoc is an exception, seeing less than
1.5x speedup.  The validator generally has a higher speedup than the
parallel miner.  This is because the parallel miner has done the hard
work of finding data conflict and produced a locking schedule for the
validator to follow.

The right-hand chart of Figure~\ref{fig:speedup} plots the speedup as the
data conflict percentage increases for fixed blocks of 200 transactions.
As data conflict increases, the miner's speedup reduces from 2x to close
to serial as many transactions touch shared data.  The validator also
starts at around 2x with no data conflict, but goes down to about 1.5x,
again benefiting from the work of the parallel miner.

Ballot's parallel
mining hovers around 1.5x speedup, suffering little from the extra
data conflict.  Data Conflict in SimpleAuction and EtherDoc, however, has an
expectedly higher impact, because each contending transaction touches
the same data.  The Mixed benchmark provides a more realistic view of
a block by combining transactions from unrelated contracts.  Even though EtherDoc
reduces parallelism under high data conflict, when mixed with other transactions, the parallel
miner can still gain a substantial speedup.

\iftr
\begin{table}
\begin{center}
\begin{tabular}{|l|r|r|r|r|r|r|r|r|}
\hline
 & \multicolumn{2}{|c|}{\vvv{SimpleAuction}} & \multicolumn{2}{|c|}{\vvv{Ballot}}& \multicolumn{2}{|c|}{\vvv{EtherDoc}} & \multicolumn{2}{|c|}{\vvv{Mixed}}\\
 & Conflict & BlockSize & Conflict & BlockSize & Conflict & BlockSize & Conflict & BlockSize \\
\hline
{\bf Miner} & 
  \SimpleAuctioncontentionMiner{}x
 & \SimpleAuctionblocksizeMiner{}x
 & \BallotcontentionMiner{}x
 & \BallotblocksizeMiner{}x
 & \EtherDoccontentionMiner{}x
 & \EtherDocblocksizeMiner{}x
 & \MixedcontentionMiner{}x
 & \MixedblocksizeMiner{}x
\\
{\bf Validator} & \SimpleAuctioncontentionValidator{}x
 & \SimpleAuctionblocksizeValidator{}x
 & \BallotcontentionValidator{}x
 & \BallotblocksizeValidator{}x
 & \EtherDoccontentionValidator{}x
 & \EtherDocblocksizeValidator{}x
 & \MixedcontentionValidator{}x
 & \MixedblocksizeValidator{}x\\
 \hline
\end{tabular}
\end{center}

\caption{The average speedups for each benchmark.}
\label{tab:speedups}
\end{table}
\fi

The average of speedups of all benchmarks is \overallMinerSpeedup{}x for the
parallel miner and \overallValidatorSpeedup{}x for the validator.
\iftr
Table~\ref{tab:speedups} shows the average speedups for each benchmark.
\else
The complete table of average speedups can be found in in
Appendix~\ref{apx:speeduptable}
\fi

\subsection{Discussion}
These results show that speculative concurrent execution speeds up
mining when threads are occupied and the data conflict rate is not too high.
Data conflicts among transactions in the same block is likely to be
infrequent over the long term.
(Miners could also choose transactions so as to reduce the likelihood of conflict,
say by including only those contracts that operate on disjoint data sets.)
Due to limited hardware,
our experiments used only three concurrent threads,
but even this modest level of concurrency showed a benefit.
Concurrent hardware has proved effective for speeding up solutions to proof-of-work puzzles,
and now similar investments could speed up smart contract execution and validation.

\section{Related Work}
The notion of smart contracts can be traced back to an article by Nick
Szabo in 1997~\cite{Szabo1997}.
Bitcoin~\cite{bitcoin} includes a scripting language whose expressive
power was limited to protect against non-terminating scripts.
Ethereum~\cite{ethereum} is perhaps the most widely used smart
contract platform,
employing a combination of a Turing-complete virtual machine protected from
non-termination by charging clients for contract running times.
Solidity~\cite{solidity} is the most popular programming language for
programming the Ethereum virtual machine.

Luu \emph{et al.}~\cite{DBLP:conf/ccs/LuuCOSH16} identify a number of
security vulnerabilities and pitfalls in the Ethereum smart contract
model.
Luu \emph{et al.}~\cite{Luu:2015:DIC:2810103.2813659} also identify
perverse incentives that cause rational miners sometimes to accept
unvalidated blocks.
Delmolino \emph{et al.}~\cite{Delmolino2016}
document common programming errors observed in smart contracts.
The Hawk~\cite{Kosba2015HawkTB} smart contract system is designed to protect the privacy of participants.

As noted, many of the speculative mechanisms introduced here were
adapted from \emph{transactional boosting}~\cite{HerlihyK2008},
a technique for transforming thread-safe linearizable objects into
highly-concurrent transactional objects.
Boosting was originally developed to enhance the concurrency provided by
software transactional memory
systems~\cite{Herlihy:2003:STM:872035.872048} by exploiting
type-specific information.
Other techniques that exploit type-specific properties to enhance
concurrency in STMs include
\emph{transactional predication}~\cite{BronsonCCOn2010}
and \emph{software transactional objects}~\cite{HermanIHTKLS2016}.

There are other techniques for deterministically reproducing a prior concurrent execution.
See Bocchino \emph{et al.}~\cite{Bocchino:2009:PPM:1855591.1855595} for a survey.

\section{Conclusion}

We have shown that one can exploit multi-core architectures to
increase smart contract processing throughput for both miners and validators.
First, miners execute a block's contracts speculatively and in parallel,
resulting in lower latency whenever the block's contracts lack data conflicts.
Miners are incentivized to include in each block an encoding of the
serializable parallel schedule that produced that block.
Validators convert that schedule into a deterministic, parallel
fork-join program that allows them to validate the block in parallel.
Even with only three threads,
a prototype implementation yields overall
speedups of \overallMinerSpeedup{}x for miners and
\overallValidatorSpeedup{}x for validators on representative
smart contracts.

Future work could include adding support for multithreading to
the Ethereum virtual machine, in much the same way as today's Java virtual machines.
Our proposal for miners only is compatible with current smart contract systems such as Ethereum,
but our overall proposal is not,
because it requires including scheduling metadata in blocks,
and incentivizing miners to publish their parallel schedules.
It may well be compatible with a future ``soft fork'' (backward
compatible change), a subject for future research.

\vfill
\pagebreak

\bibliographystyle{abbrv}
\bibliography{blockchain}

\begin{thebibliography}{10}

\bibitem{BlumofeJKLRZ1995}
R.~D. Blumofe, C.~F. Joerg, B.~C. Kuszmaul, C.~E. Leiserson, K.~H. Randall, and
  Y.~Zhou.
\newblock Cilk: An efficient multithreaded runtime system.
\newblock In {\em Proceedings of the Fifth ACM SIGPLAN Symposium on Principles
  and Practice of Parallel Programming}, PPOPP '95, pages 207--216, New York,
  NY, USA, 1995. ACM.

\bibitem{Bocchino:2009:PPM:1855591.1855595}
R.~L. Bocchino, Jr., V.~S. Adve, S.~V. Adve, and M.~Snir.
\newblock Parallel programming must be deterministic by default.
\newblock In {\em Proceedings of the First USENIX Conference on Hot Topics in
  Parallelism}, HotPar'09, pages 4--4, Berkeley, CA, USA, 2009. USENIX
  Association.

\bibitem{BronsonCCOn2010}
N.~G. Bronson, J.~Casper, H.~Chafi, and K.~Olukotun.
\newblock Transactional predication: High-performance concurrent sets and maps
  for stm.
\newblock In {\em Proceedings of the 29th ACM SIGACT-SIGOPS Symposium on
  Principles of Distributed Computing}, PODC '10, pages 6--15, New York, NY,
  USA, 2010. ACM.

\bibitem{theDao}
DAO.
\newblock Thedao smart contract.
\newblock Retrieved 8 February 2017.

\bibitem{Delmolino2016}
K.~Delmolino, M.~Arnett, A.~Kosba, A.~Miller, and E.~Shi.
\newblock {\em Step by Step Towards Creating a Safe Smart Contract: Lessons and
  Insights from a Cryptocurrency Lab}, pages 79--94.
\newblock Springer Berlin Heidelberg, Berlin, Heidelberg, 2016.

\bibitem{ethereum}
Ethereum.
\newblock \url{https://github.com/ethereum/}.

\bibitem{ethereumdesign}
{Ethereum design Rationale}.
\newblock
  \url{http://github.com/ethereum/wiki/wiki/Design-Rationale\#gas-and-fees}.

\bibitem{opacity}
R.~Guerraoui and M.~Kapalka.
\newblock On the correctness of transactional memory.
\newblock In {\em Proceedings of the 13th ACM SIGPLAN Symposium on Principles
  and practice of parallel programming (PPoPP'08)}, pages 175--184, New York,
  NY, USA, 2008. ACM.

\bibitem{HerlihyK2008}
M.~Herlihy and E.~Koskinen.
\newblock Transactional boosting: A methodology for highly-concurrent
  transactional objects.
\newblock In {\em Proceedings of the 13th ACM SIGPLAN Symposium on Principles
  and Practice of Parallel Programming}, PPoPP '08, pages 207--216, New York,
  NY, USA, 2008. ACM.

\bibitem{Herlihy:2003:STM:872035.872048}
M.~Herlihy, V.~Luchangco, M.~Moir, and W.~N. Scherer, III.
\newblock Software transactional memory for dynamic-sized data structures.
\newblock In {\em Proceedings of the twenty-second annual symposium on
  Principles of distributed computing}, PODC '03, pages 92--101, New York, NY,
  USA, 2003. ACM.

\bibitem{HermanIHTKLS2016}
N.~Herman, J.~P. Inala, Y.~Huang, L.~Tsai, E.~Kohler, B.~Liskov, and L.~Shrira.
\newblock Type-aware transactions for faster concurrent code.
\newblock In {\em Proceedings of the Eleventh European Conference on Computer
  Systems}, EuroSys '16, pages 31:1--31:16, New York, NY, USA, 2016. ACM.

\bibitem{Kosba2015HawkTB}
A.~E. Kosba, A.~Miller, E.~Shi, Z.~Wen, and C.~Papamanthou.
\newblock Hawk: The blockchain model of cryptography and privacy-preserving
  smart contracts.
\newblock In {\em IEEE Symposium on Security and Privacy}, 2015.

\bibitem{KP:PLDI2015}
E.~Koskinen and M.~J. Parkinson.
\newblock The push/pull model of transactions.
\newblock In {\em Proceedings of the 36th ACM SIGPLAN Conference on Programming
  Language Design and Implementation (PLDI'15), Portland, OR, USA}. ACM, 2015.

\bibitem{KPH:POPL2010}
E.~Koskinen, M.~J. Parkinson, and M.~Herlihy.
\newblock Coarse-grained transactions.
\newblock In {\em Proceedings of the 37th ACM SIGPLAN-SIGACT Symposium on
  Principles of Programming Languages (POPL'10)}, pages 19--30. ACM, 2010.

\bibitem{DBLP:conf/ccs/LuuCOSH16}
L.~Luu, D.~Chu, H.~Olickel, P.~Saxena, and A.~Hobor.
\newblock Making smart contracts smarter.
\newblock In {\em Proceedings of the 2016 {ACM} {SIGSAC} Conference on Computer
  and Communications Security, Vienna, Austria, October 24-28, 2016}, pages
  254--269, 2016.

\bibitem{Luu:2015:DIC:2810103.2813659}
L.~Luu, J.~Teutsch, R.~Kulkarni, and P.~Saxena.
\newblock Demystifying incentives in the consensus computer.
\newblock In {\em Proceedings of the 22Nd ACM SIGSAC Conference on Computer and
  Communications Security}, CCS '15, pages 706--719, New York, NY, USA, 2015.
  ACM.

\bibitem{bitcoin}
S.~Nakamoto.
\newblock Bitcoin: A peer-to-peer electronic cash system.
\newblock May 2009.

\bibitem{scalastm}
{Scala STM Expert Group.}
\newblock Scalastm. web.
\newblock Retrieved from \url{http://nbronson.github.com/scala-stm/}, 20
  November 2011.

\bibitem{solidity}
{Solidity documentation}.
\newblock \url{http://solidity.readthedocs.io/en/latest/index.html}.

\bibitem{solidityexamples}
{Solidity documentation: Solidity by example}.
\newblock
  \url{http://solidity.readthedocs.io/en/develop/solidity-by-example.html}.

\bibitem{Szabo1997}
N.~Szabo.
\newblock Formalizing and securing relationships on public networks.
\newblock {\em First Monday}, 2(9), 1997.

\bibitem{wood}
G.~Wood.
\newblock Ethereum: A secure decentralised generalised transaction ledger.

\end{thebibliography}

\pagebreak
\appendix
\section{Example Contract: Ballot}
\label{apx:contract}

\begin{lstlisting}
pragma solidity ^0.4.0;
/// @title Voting with delegation.
contract Ballot {
    // This declares a new complex type which will
    // be used for variables later.
    // It will represent a single voter.
    struct Voter {
        uint weight; // weight is accumulated by delegation
        bool voted;  // if true, that person already voted
        address delegate; // person delegated to
        uint vote;   // index of the voted proposal
    }
    // This is a type for a single proposal.
    struct Proposal
    {
        bytes32 name;   // short name (up to 32 bytes)
        uint voteCount; // number of accumulated votes
    }
    address public chairperson;
    // This declares a state variable that
    // stores a `Voter` struct for each possible address.
    mapping(address => Voter) public voters;
    // A dynamically-sized array of `Proposal` structs.
    Proposal[] public proposals;
    /// Create a new ballot to choose one of `proposalNames`.
    function Ballot(bytes32[] proposalNames) {
        chairperson = msg.sender;
        voters[chairperson].weight = 1;
        // For each of the provided proposal names,
        // create a new proposal object and add it
        // to the end of the array.
        for (uint i = 0; i < proposalNames.length; i++) {
            // `Proposal({...})` creates a temporary
            // Proposal object and `proposals.push(...)`
            // appends it to the end of `proposals`.
            proposals.push(Proposal({
                name: proposalNames[i],
                voteCount: 0
            }));
        }
    }
    // Give `voter` the right to vote on this ballot.
    // May only be called by `chairperson`.
    function giveRightToVote(address voter) {
        if (msg.sender != chairperson || voters[voter].voted) {
            // `throw` terminates and reverts all changes to
            // the state and to Ether balances. It is often
            // a good idea to use this if functions are
            // called incorrectly. But watch out, this
            // will also consume all provided gas.
            throw;
        }
        voters[voter].weight = 1;
    }
    /// Delegate your vote to the voter `to`.
    function delegate(address to) {
        // assigns reference
        Voter sender = voters[msg.sender];
        if (sender.voted)
            throw;
        // Forward the delegation as long as
        // `to` also delegated.
        // In general, such loops are very dangerous,
        // because if they run too long, they might
        // need more gas than is available in a block.
        // In this case, the delegation will not be executed,
        // but in other situations, such loops might
        // cause a contract to get "stuck" completely.
        while (
            voters[to].delegate != address(0) &&
            voters[to].delegate != msg.sender
        ) {
            to = voters[to].delegate;
        }
        // We found a loop in the delegation, not allowed.
        if (to == msg.sender) {
            throw;
        }
        // Since `sender` is a reference, this
        // modifies `voters[msg.sender].voted`
        sender.voted = true;
        sender.delegate = to;
        Voter delegate = voters[to];
        if (delegate.voted) {
            // If the delegate already voted,
            // directly add to the number of votes
            proposals[delegate.vote].voteCount += sender.weight;
        } else {
            // If the delegate did not vote yet,
            // add to her weight.
            delegate.weight += sender.weight;
        }
    }
    /// Give your vote (including votes delegated to you)
    /// to proposal `proposals[proposal].name`.
    function vote(uint proposal) {
        Voter sender = voters[msg.sender];
        if (sender.voted)
            throw;
        sender.voted = true;
        sender.vote = proposal;
        // If `proposal` is out of the range of the array,
        // this will throw automatically and revert all
        // changes.
        proposals[proposal].voteCount += sender.weight;
    }
    /// @dev Computes the winning proposal taking all
    /// previous votes into account.
    function winningProposal() constant
            returns (uint winningProposal)
    {
        uint winningVoteCount = 0;
        for (uint p = 0; p < proposals.length; p++) {
            if (proposals[p].voteCount > winningVoteCount) {
                winningVoteCount = proposals[p].voteCount;
                winningProposal = p;
            }
        }
    }
    // Calls winningProposal() function to get the index
    // of the winner contained in the proposals array and then
    // returns the name of the winner
    function winnerName() constant
            returns (bytes32 winnerName)
    {
        winnerName = proposals[winningProposal()].name;
    }
}
\end{lstlisting}

\pagebreak
\section{Mean and Standard Deviation of Benchmark Running Times}
\label{apx:dev}

\begin{center}
\begin{tabular}{cc}
\includegraphics[width=3.2in]{\benchstampt{Ballot-blocksize-raw.pdf}} &
\includegraphics[width=3.2in]{\benchstampt{Ballot-contention-raw.pdf}}\\
\includegraphics[width=3.2in]{\benchstampt{SimpleAuction-blocksize-raw.pdf}} &
\includegraphics[width=3.2in]{\benchstampt{SimpleAuction-contention-raw.pdf}}\\
\includegraphics[width=3.2in]{\benchstampt{EtherDoc-blocksize-raw.pdf}} &
\includegraphics[width=3.2in]{\benchstampt{EtherDoc-contention-raw.pdf}}\\
\includegraphics[width=3.2in]{\benchstampt{Mixed-blocksize-raw.pdf}} &
\includegraphics[width=3.2in]{\benchstampt{Mixed-contention-raw.pdf}}\\
\end{tabular}
\end{center}

\iftr
\else
\pagebreak
\section{Average Speedups for All Benchmarks}
\label{apx:speeduptable}

\begin{center}
\begin{tabular}{|l|r|r|r|r|r|r|r|r|}
\hline
 & \multicolumn{2}{|c|}{\vvv{SimpleAuction}} & \multicolumn{2}{|c|}{\vvv{Ballot}}& \multicolumn{2}{|c|}{\vvv{EtherDoc}} & \multicolumn{2}{|c|}{\vvv{Mixed}}\\
 & Conflict & BlockSize & Conflict & BlockSize & Conflict & BlockSize & Conflict & BlockSize \\
\hline
{\bf Miner} & 
  \SimpleAuctioncontentionMiner{}x
 & \SimpleAuctionblocksizeMiner{}x
 & \BallotcontentionMiner{}x
 & \BallotblocksizeMiner{}x
 & \EtherDoccontentionMiner{}x
 & \EtherDocblocksizeMiner{}x
 & \MixedcontentionMiner{}x
 & \MixedblocksizeMiner{}x
\\
{\bf Validator} & \SimpleAuctioncontentionValidator{}x
 & \SimpleAuctionblocksizeValidator{}x
 & \BallotcontentionValidator{}x
 & \BallotblocksizeValidator{}x
 & \EtherDoccontentionValidator{}x
 & \EtherDocblocksizeValidator{}x
 & \MixedcontentionValidator{}x
 & \MixedblocksizeValidator{}x\\
 \hline
\end{tabular}
\end{center}

\fi

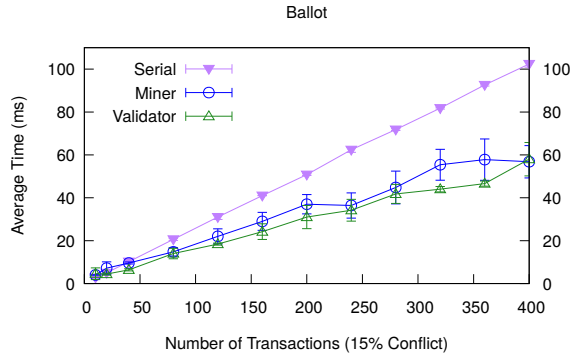
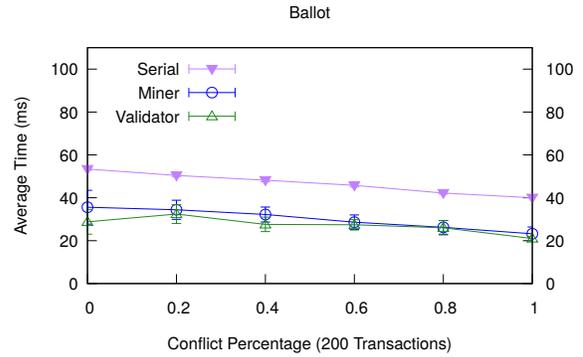
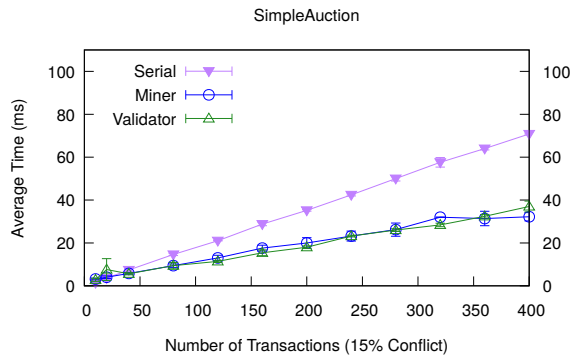
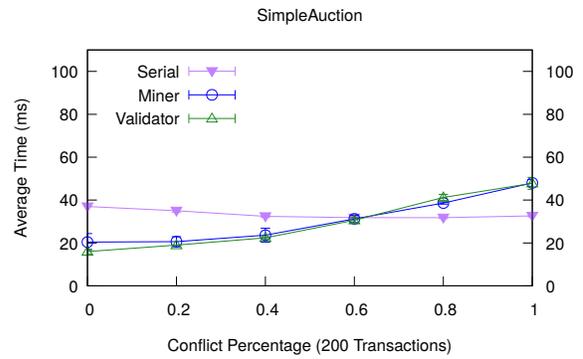
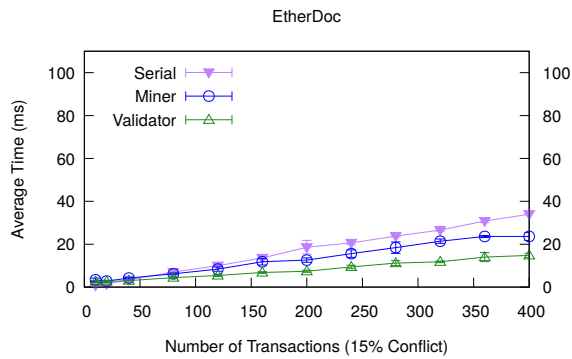
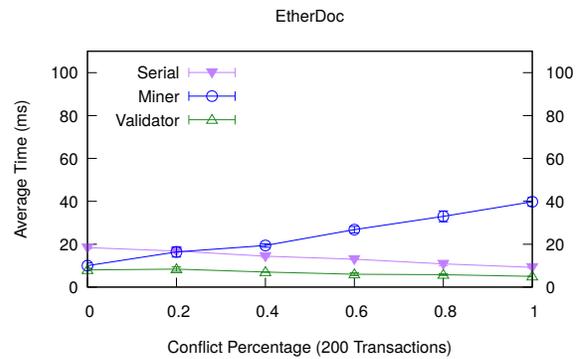
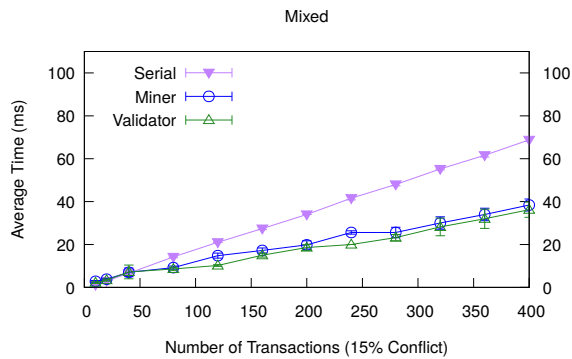
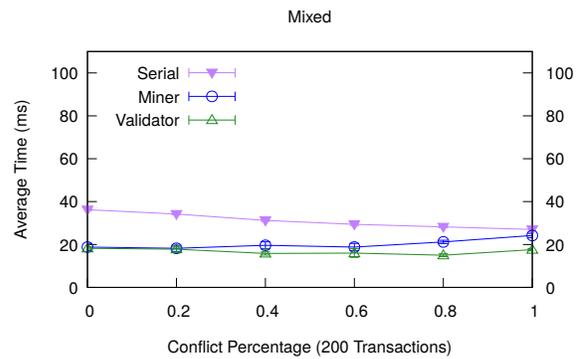
\end{document}